\documentclass[prl,twocolumn,amsmath,amssymb]{revtex4} 
\usepackage{graphicx} 
\usepackage{amsmath}
\usepackage{amsfonts,amsbsy}
\usepackage{amssymb}

\usepackage{subfig}
\def\gsim{ \,\, \vcenter{\hbox{$\buildrel{\displaystyle >}\over\sim$}}
 \,\,}

\def\be{\begin{equation}}
\def\ee{\end{equation}}
\def\bea{\begin{eqnarray}}
\def\eea{\end{eqnarray}}
\def\tr{{\rm tr}\,}
\begin{document}

\title{\bf Spatial Wilson loops in the classical field of
  high-energy heavy-ion collisions}

\author{Elena Petreska$^{a,b}$}
\affiliation{
$^a$ Department of Natural Sciences, Baruch College, CUNY,
17 Lexington Avenue, New York, NY 10010, USA\\
$^b$ The Graduate School and University Center, The City
  University of New York, 365 Fifth Avenue, New York, NY 10016, USA
}

\begin{abstract}

It has been previously shown numerically that the expectation value of
the magnetic Wilson loop at the initial time of a heavy-ion collision
exhibits area law scaling. This was obtained for a classical
non-Abelian gauge field in the forward light cone and for loops of
area $A\gsim 2/Q_s^2$. Here, we present an analytic calculation of the
spatial Wilson loop in the classical field of a collision
within perturbation theory. It corresponds to a diagram with two
sources, for both projectile and target, whose field is evaluated at
second order in the gauge potential. The leading non-trivial
contribution to the magnetic loop in perturbation theory is
proportional to the square of its area.

\end{abstract}

\maketitle

In high-energy collisions the target and the projectile are
represented as Lorentz-contracted sheets of valence color charges
moving along recoilless trajectories along the light cone. These
charges act as sources of a purely transverse gluon field that has a
small fraction $x$ of the total longitudinal momentum of the
nucleus. Since the charge density $\rho (\bf x) $ is large, the
sources belong to a higher dimensional representation of the color
algebra and the gauge field they emit can be computed classically
\cite{MV}. After the two sheets of color charge have passed through
each other, longitudinal chromo-electric and chromo-magnetic fields
are produced \cite{Fries:2006pv_TL_LM_KKV}. The fluctuations of the
chromo-magnetic flux may be viewed as uncorrelated vortices with a
typical radius $\sim 0.8/Q_s$ \cite{DNP}. $Q_s$ denotes the saturation
momentum which is the scale where the gluon field exhibits non-linear
dynamics \cite{Saturation}. The effective area law behavior of the
magnetic flux which indicated vortex structure was obtained in
ref.~\cite{DNP} numerically. Here we provide some analytic insight
within perturbation theory and compare it to the lattice
computation. Since the perturbative expansion of the magnetic flux
applies only to small loops ($A\ll 1/Q_s^2$) it may of course deviate
from the lattice result obtained for large loops. Indeed, the latter resums
screening corrections to the magnetic field~\cite{DumitruFujiiNara}.

The gluon field of the target and the projectile is obtained by
solving the classical Yang-Mills equations of motion. Before the
collision, the solution corresponds to a non-Abelian analogue of the
Weizs{\"a}cker-Williams field. In light-cone gauge its form is:
\be \label{eq:alphai} 
\alpha^i_m = \frac{i}{g} \, U_m \, \partial^i
U_m^\dagger~~~~,~~~~ \partial^i \alpha^i_m = g \rho_m~.
\ee 
The subscript $m$, with values 1 and 2, denotes the projectile and the
target respectively. Introducing the gauge potential as
\be 
\Phi_m=-\frac{g}{\nabla_\perp^2}\rho_m~,
\ee
the solution to (\ref{eq:alphai}) can be written as \cite{Kovner:1995ja}:
\be \label{eq:alphaUU}
\alpha_m^i = \frac{i}{g}e ^{-ig\Phi_m}\partial^i e ^{ig\Phi_m}~.
\ee
The Yang-Mills equations for scattering of two ultra-relativistic
nuclei with appropriate boundary conditions on the light cone give the
classical field after the collision \cite{Kovner:1995ja}. At proper
time $\tau=\sqrt{t^2-z^2}=0$, the resulting field is a sum
of two pure gauge fields:
\be 
A^i=\alpha_1^i + \alpha_2^i~.
\ee
The sum of two pure gauges is not a pure gauge and
strong longitudinal chromo-electric and chromo-magnetic fields are
produced in the collision \cite{Fries:2006pv_TL_LM_KKV}:
\bea
E_z&=&ig[\alpha_1^i,\alpha_2^i]~,\nonumber \\
B_z&=&ig\epsilon^{ij}[\alpha_1^i,\alpha_2^j]~~~,~~~(i,j=1,2)~. 
\eea
$\epsilon^{ij}$ is the antisymmetric tensor. The transverse components of the field strength are zero.

The non-Abelian Wilson loop operator is defined as a path order
exponential of the gauge field:
\bea \label{eq:M_def}
M(R) &=& {\cal P}\exp\left(ig \oint  dx^iA^i\right) \nonumber\\
&=& {\cal P}\exp\left[ig \oint  dx^i\left(
  \alpha_1^i+\alpha_2^i\right)\right]~,
\eea
with $R$ the radius of the loop. Note that $M(R)\equiv 1\!\! 1$ if
evaluated in the field of a single nucleus ($\alpha_1^i$ or
$\alpha_2^i$) as those are pure gauges. In \cite{DNP} it was shown
that the expectation value of the magnetic Wilson loop in the field
$A^i$ produced in a collision of two nuclei is proportional to the
exponent of the area $A$ of the loop:
\be \label{eq:W_M_def}
W_M(R) = \frac{1}{N_c} \left< \tr M(R)\right> \sim \exp\left(-\sigma_M
A\right)~.
\ee
Here, $\sigma_M$ is the magnetic string tension. For the SU(2) gauge
group its value was estimated to
be $\sigma_M \simeq 0.12Q_s^2$. The result (\ref{eq:W_M_def}) was
obtained for areas $A\gsim 2/Q_s^2$. It indicates that the
structure of the chromo-magnetic flux at such scales corresponds to
uncorrelated vortex fluctuations.

The expectation value in~(\ref{eq:W_M_def}) refers to averaging over
the color charge distributions in each nucleus. For large nuclei the
color sources are treated as random variables with Gaussian
probability distribution. Physical observables are then averaged with
a Gaussian (McLerran-Venugopalan) action:
\be  \label{eq:S2}
S_{\rm eff}[\rho^a] = \frac{1}{2}
 \int d^2\bf{x} \left[\frac{\rho_1^a({\bf x}) \rho_1^a({\bf x})}{\mu^2_1}
  +
\frac{\rho_2^a({\bf x}) \rho_2^a({\bf x})}{\mu^2_2}\right]
~,
\ee 
where $\mu^2$ is the color charge squared per unit area, related to
the saturation scale via $Q_s^2 \sim g^4 \mu^2$.

To obtain $W_M(R)$ we need to determine the deviation of $A^i$ from a
pure gauge. From the Baker-Campbell-Hausdorff formula~\cite{DNP}
\be \label{eq:BCH}
W_M(R) \simeq \frac{1}{N_c}  \left<\tr  \exp
\left(-\frac{1}{2}\left[X_1,X_2\right]\right)\right> \simeq 1-\frac{1}{2N_c}\left<g^2h^2\right> ~,
\ee
where: 
\be 
g^2h^2=\frac{1}{8}f^{abc}f^{\bar{a}\bar{b}c}X_1^aX_1^{\bar{a}}X_2^bX_2^{\bar{b}}~. 
\ee
with:
\be 
X_m = ig \oint dx^i \alpha_m^{ai}t^a~. 
\ee
$f^{abc}$ are the structure constants of the special unitary group and
$h^2$ corresponds to a four gluon vertex of the fields.

\begin{figure}
\subfloat[]{\label{fig:1}\includegraphics[width=0.105\textwidth]{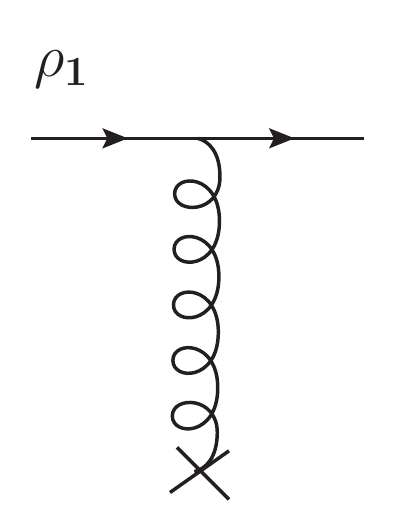}}\qquad
\subfloat[]{\label{fig:2}\includegraphics[width=0.341\textwidth]{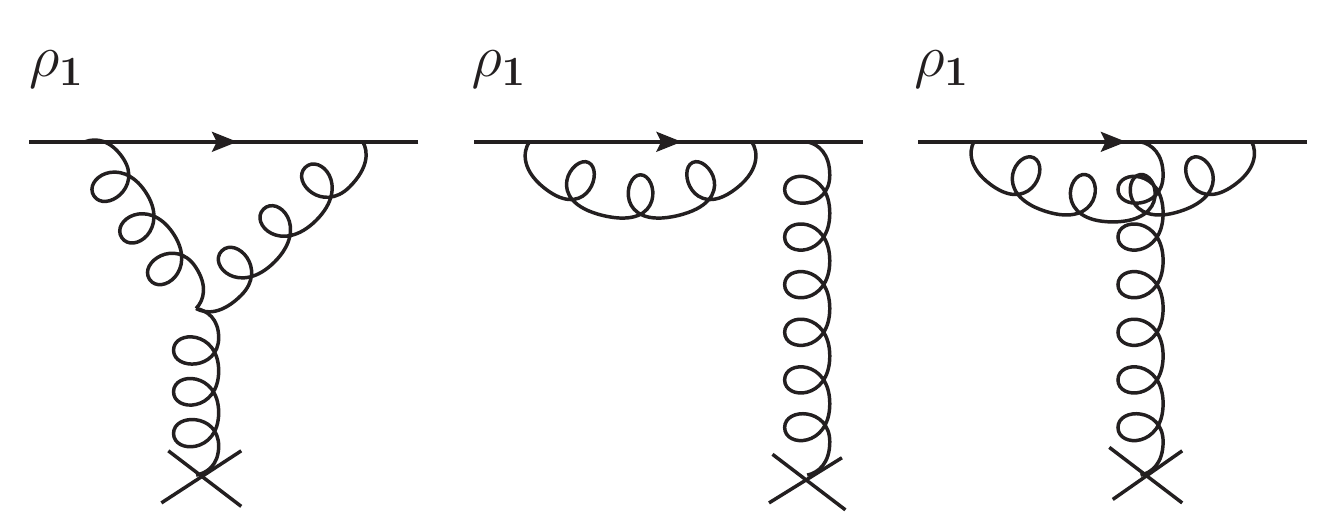}}
\caption{\label{fig:alpha_sngl}Diagram representation of the classical
  gauge field of a single nucleus to first (\ref{fig:1}) and second
  order (\ref{fig:2}) \cite{Kovchegov}.}
\end{figure}

To calculate this expectation value analytically, we expand the fields
$\alpha^i$ from eq.~(\ref{eq:alphaUU}) perturbatively in terms of the
coupling constant:
\bea \label{eq:alpha_pert}
\alpha_m^i = &-& \partial^i\Phi_m +
\frac{ig}{2}\left(\delta^{ij}-\partial^i\frac{1}{\nabla_\perp^2}\partial^j\right)
\left[\Phi_m,\partial^j  \Phi_m\right] \nonumber \\
&+&\mathcal{O}\left(\Phi_m^3\right)~. 
\eea
The loop integral over the first term in~(\ref{eq:alpha_pert})
vanishes. Therefore, the leading term in the expansion of the field
$\alpha^i$ in terms of the gauge potential $\Phi$ does not contribute
to the Wilson loop $W_M$ and it is required to go to quadratic order:
$\alpha^{i,a} \sim gf^{abc}\Phi^b\partial^i\Phi^c$:  
\be 
X_m=-\frac{g^2}{2} \oint dx^i ~\left[\Phi_m , \partial^i \Phi_m\right] ~. 
\ee
The Feynman
diagram representation of the expansion of the field $\alpha^i$ is
given in fig.~\ref{fig:alpha_sngl}. At this order the field is still
dominated by classical diagrams~\cite{Kovchegov}.

\begin{figure}
\subfloat[]{\label{fig:s1}\includegraphics[width=0.18\textwidth]{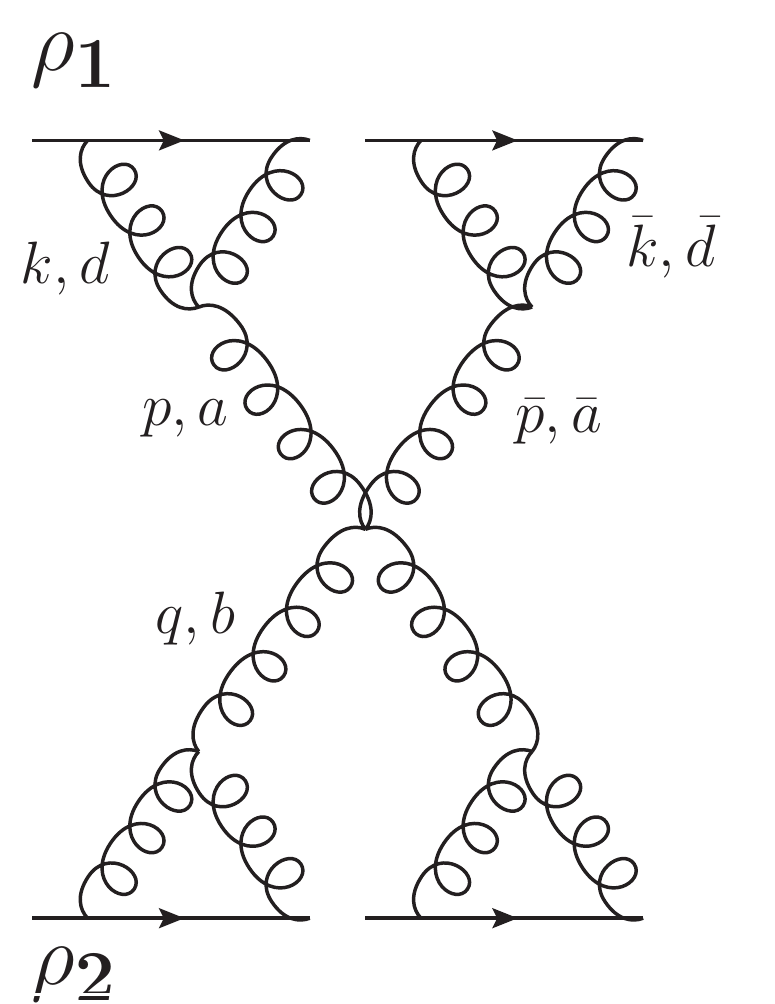}}\qquad
\subfloat[]{\label{fig:s2}\includegraphics[width=0.155\textwidth]{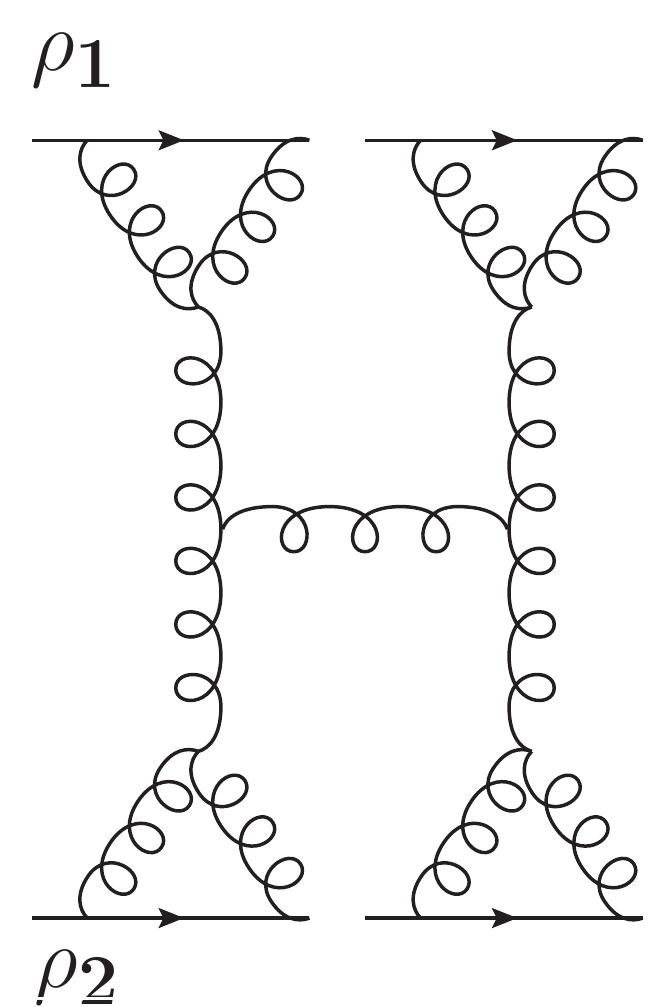}}
\caption{\label{fig:s}Classical (\ref{fig:s1}) and quantum (\ref{fig:s2}) contributions to the expectation value of the Wilson loop.}
\end{figure}

The expectation value $\left<h^2\right>$ that enters in the expression
for the magnetic loop involves the fields of both nuclei.
The corresponding classical diagram is shown in
fig.~\ref{fig:s1}. A quantum correction at the same order 
is given in fig.~\ref{fig:s2}. We defer
an analysis of quantum corrections to future work since the primary
goal of this paper is to provide a point of comparison to resummed
classical lattice gauge computations of the loop~\cite{DNP} which
consider strong fields. However, our present analysis should not be
taken to imply that small loops in weak fields can be obtained in the
classical approximation.

The final result we obtain for the expectation value of the magnetic Wilson loop for classical fields $\alpha^i$ to second order in the gauge potential is:
\be \label{eq:Final_W_M}
W_M(R) \simeq 1-\frac{\pi^2 N_c^6}{64(N_c^2-1)^3}\frac{Q_{s1}^4Q_{s2}^4}{\Lambda^4}A^2~.
\ee
Details of the calculation are given in the appendix. $Q_{s1}$ and
$Q_{s2}$ are the saturation scales of the projectile and the target,
respectively. They are determined by the variance of the color charge
distribution. We use the relation:
\be \label{eq:Qs}
Q_{s}^2=\frac{g^4C_F}{2\pi}\mu^2~, 
\ee
where:
\be 
C_F=\frac{N_c^2-1}{2N_c}~.
\ee
The cut-off $\Lambda$ regulates the infrared divergence of the
integrals over the gluon momentum $k$ shown in diagram
\ref{fig:s1}. It sets the mass scale for the gluon propagator.

From a fit to the lattice data for the Wilson loop for small areas, it
was estimated that $W_M(R) \simeq 1-2\left(AQ_s^2\right)^2$
\cite{DNP}. By comparing this expression to the result
(\ref{eq:Final_W_M}) for $SU(2)$ we can extract
$Q_s^4/\Lambda^4\approx 5.477$.

We have calculated the expectation value of the magnetic loop with a
Gaussian action. For a finite nuclear thickness, higher order corrections
in the charge density of cubic \cite{JeonVenugopalan} and quartic
\cite{DJP} order arise. As shown in the appendix, the calculation 
consists of averaging four-point
functions and therefore the cubic part of the
effective action does not contribute. On the other hand, one would
expect the fourth order term to bring a correction to the four-point
function. However, the correction to the Wilson loop from the quartic
action vanishes because of its vanishing color factor (see appendix).

We now turn to a discussion of the final result
(\ref{eq:Final_W_M}). The perturbative result for the 
expectation value of the magnetic Wilson loop
gives a leading non-trivial contribution that is proportional to the
square of the area. 
A term proportional to the area of the loop would involve single powers of
the target's and projectile's saturation scales: $\sim
A~Q_{s1}Q_{s2}$~\cite{DNP}. However, Gaussian contractions can only give
powers of $Q_{s1}^2$ and $Q_{s2}^2$:
\be
\langle \rho^a_m({\bf x})\, \rho^b_m({\bf y})\rangle =\mu_m^2
\delta^{ab} \delta({\bf x}-{\bf y})\sim Q_{s_m}^2~,
\ee
and therefore a contribution $\sim A^2$. Area law scaling of the
Wilson loop presumably requires resummation of screening effects and
of condensation.

%---------------------------------------------------------------------------

%---------------------------------------------------------------------
\vspace*{1cm}
\begin{acknowledgments}
I thank A.~Dumitru for helpful discussions.
Support by the DOE
Office of Nuclear Physics through Grant No.\ DE-FG02-09ER41620; and
from The City University of New York through the PSC-CUNY Research
Award Program, grant 66514-0044 is gratefully acknowledged.

\end{acknowledgments}

%---------------------------------------------------------------------

\section{Appendix}

In the appendix we list some steps of the calculation leading to the final result (\ref{eq:Final_W_M}). To calculate the expectation value of the Wilson loop, we need the average:
\be 
\left<g^2h^2\right>=\frac{1}{8}f^{abc}f^{\bar{a}\bar{b}c}\left<X_1^aX_1^{\bar{a}}\right>_{\rho_1}\left<X_2^bX_2^{\bar{b}}\right>_
{\rho_2}~. 
\ee
Using the second term in the expression for the fields $\alpha_m$ in
eq.~(\ref{eq:alpha_pert}) we have:
\be 
X_m^a=-\frac{ig^2}{2}f^{ade} \oint dx^i ~\Phi_m^d \partial^i \Phi_m^e ~, 
\ee
or, in momentum space:
\bea
&&X_m^a = -\frac{ig^2}{2(2\pi)^3}f^{ade}R \times \nonumber \\
&&\int d^2{\bf k} d^2{\bf p}|{\bf k}| J_1(R|{\bf p}|) ~\sin(\alpha-\theta)\Phi_m^d({\bf k}) \Phi_m^e({\bf p}-{\bf k})~.~~~ 
\eea
In the above expression, $R$ is the radius of the loop, $k$ and $p$
are the momenta of the gluons shown in fig.~\ref{fig:s1}, and
$\alpha$ and $\theta$ are their corresponding azimuthal
angles. $J_1(R|{\bf p}|)$ is a Bessel function of the first
kind. Then:
\bea \label{eq:4phi}
&&\left<X_m^aX_m^{\bar a}\right>_{\rho_m}=-\frac{g^4}{4(2\pi)^6}f^{ade}f^{\bar a \bar d \bar e}R^2 \times \nonumber \\
&&\int d^2{\bf k}~ d^2 {\bf p}~ d^2 {\bf \bar{k}}~ d^2 {\bf \bar{p}} ~|{\bf k}||{\bf \bar{k}}|~J_1(R|{\bf p}|) J_1(R|{\bf \bar{p}}|)~\times \nonumber \\
&&\sin(\alpha-\theta)\sin(\bar{\alpha}-\bar{\theta})\times \nonumber \\
&&\left<\Phi_m^d({\bf k}) ~\Phi_m^e({\bf p}-{\bf k})~\Phi_m^{\bar d}({\bf \bar{k}}) ~\Phi_m^{\bar e}({\bf \bar{p}}-{\bf \bar{k}})\right>_{\rho_m}~. 
\eea
The gauge potential and the two-point function in momentum space are
\bea 
 \Phi^a({\bf k})=-\frac{g}{k^2}\rho^a(\bf{k})  \label{eq:Phi^a}~~~~\text{and}~~~~~~~~~~~~~~ \\
 \langle \rho^a({\bf k})\, \rho^b({\bf p})\rangle =\mu^2 \delta^{ab} (2 \pi)^2\delta({\bf k}+{\bf p})~.\label{eq:rhorho} 
\eea 
The four point-function in (\ref{eq:4phi}) receives a contribution
from the fourth order term in the extended Gaussian action \cite{DJP}:
\bea
S_Q[\rho] = \int d^2 \bf{x} \left[
\frac{\rho^a({\bf x})\rho^a({\bf x})}{2\mu^2}
+ \frac{\rho^a({\bf x})\rho^a({\bf x})
\rho^b({\bf x})\rho^b(\bf{x})}{\kappa_4}\right]~.
\label{eq:Squartic}
\eea
The correction due to the $\rho^4$ operator is:
\be 
-32\pi^2\frac{\mu^8}{\kappa_4}\left(\delta^{de}\delta^{\bar d \bar e}+\delta^{d \bar d}\delta^{e \bar e}+\delta^{d \bar e}\delta^{e \bar d}\right)\delta\left(\bf{p}+\bf{\bar p}\right)~.
\ee
But, the total color factor of this correction to the expectation value $\left<X_m^aX_m^{\bar a}\right>_{\rho_m}$ in (\ref{eq:4phi}) is equal to zero:
\be 
f^{ade}f^{\bar a \bar d \bar e}\left(\delta^{de}\delta^{\bar d \bar e}+\delta^{d \bar d}\delta^{e \bar e}+\delta^{d \bar e}\delta^{e \bar d}\right)=0~,
\ee
and does not bring a modification to the expectation value of the Wilson loop.

With the Gaussian contractions (\ref{eq:rhorho}) the expectation value
$\left<X_m^aX_m^{\bar a}\right>_{\rho_m}$ becomes:
\bea \label{delta}
&&\left<X_m^aX_m^{\bar a}\right>_{\rho_m}=-\frac{g^8\mu_m^4 }{16\pi^2}f^{ade}f^{\bar a d e} R^2 \times\nonumber \\
&&\int d^2{\bf k}~ d^2 {\bf p}~ d^2 {\bf \bar{k}}~ d^2 {\bf \bar{p}} ~\frac{J_1(R|{\bf p}|) J_1(R|{\bf \bar{p}}|)}
{|{\bf k}||{\bf \bar{k}}|({\bf p}-{\bf k})^2({\bf \bar{p}}-{\bf \bar{k}})^2}~\times \nonumber \\ 
&&\sin(\alpha-\theta)\sin(\bar{\alpha}-\bar{\theta})\times \\
&&\left[\delta({\bf k}+{\bf \bar{k}})\delta({\bf p}-{\bf k}+{\bf \bar{p}}-{\bf \bar{k}})-\delta({\bf k}+{\bf \bar{p}}-{\bf \bar{k}})\delta({\bf p}-{\bf k}+{\bf \bar{k}})\right]~.\nonumber
\eea
After performing two of the integrals using the delta functions in
(\ref{delta}), we get:
\be 
\left<X_m^aX_m^{\bar a}\right>_{\rho_m}=\frac{g^8\mu_m^4}{8}N_c\delta^{a \bar a}R^2 
\int \frac{dk}{k^3}\int dp \frac{J_1^2(R|{\bf p}|)}{|{\bf p}|}~. 
\ee
The integral over the momentum $p$ is convergent and equal to $1/2$. The integral over $k$ is infrared divergent and we introduce a cut-off $\Lambda$ to regulate this divergence: 
\be 
\int_\Lambda^\infty \frac{dk}{k^3}=\frac{1}{2\Lambda^2}~.
\ee
So, finally:
\be 
\left<X_m^aX_m^{\bar a}\right>_{\rho_m}=\frac{g^8\mu_m^4}{32\Lambda^2}N_c\delta^{a \bar a}R^2~.  \label{eq:<XX>}
\ee
The cut-off $\Lambda$ can be thought of as due to screening of the
gauge potential. Introducing screened propagators in (\ref{eq:Phi^a}):
\be 
\Phi^a({\bf k})=-\frac{g}{k^2+m^2}\rho^a(\bf{k})~,
\ee
reproduces the result (\ref{eq:<XX>}) with $\Lambda^2$ replaced by
$m^2$. A self-consistent resummation of screening effects is beyond
the purpose of the present analysis.

In terms of the saturation scale (\ref{eq:Qs}) the final result is:
\be 
\left<g^2h^2\right>=\frac{\pi^2N_c^7}{32\left(N_c^2-1\right)^3}\frac{Q_{s1}^4Q_{s2}^4}{\Lambda^4}A^2~, 
\ee
and
\be
W_M(R) \simeq 1-\frac{\pi^2 N_c^6}{64(N_c^2-1)^3}\frac{Q_{s1}^4Q_{s2}^4}{\Lambda^4}A^2~.
\ee

\end{document}